\documentstyle[epsf,12pt]{article}                                        
\newcommand{\lsim}{\stackrel{<}{\sim}} % <~

\newcommand\be{\begin{equation}}
\newcommand\ee{\end{equation}}
\newcommand\bea{\begin{eqnarray}}
\newcommand\eea{\end{eqnarray}}

\newcommand{\phd}{\phi^{\dagger}}

\title{                                                                         
{\vspace{-2cm} \normalsize                                                      
\hfill \parbox{40mm}{DESY 97-020}\\                                      
\hfill \parbox{40mm}{CERN-TH/97-21  }}\\[25mm]                                      
A Polynomial Hybrid Monte Carlo Algorithm \\[8mm]}                     
\author{                                                                        
Roberto Frezzotti\\
Deutsches Elektronen-Synchrotron DESY, \\                                       
Notkestr.\,85, D-22603 Hamburg, Germany \\
\vspace{1cm} \\                                       
Karl Jansen\\
CERN Theory Division \\
CH-1211 Gen\`eve 23, Switzerland                                         
 }                                                                               
\begin{document}                                                                
%\addtolength{\baselineskip}{+1.0\baselineskip}
\maketitle                                                                      
                                                                                
\begin{abstract} \normalsize                                                    
We present a simulation algorithm for dynamical fermions that
combines the multiboson technique with the Hybrid Monte Carlo
algorithm. We find that the algorithm gives a substantial gain over the
standard methods in practical simulations. We point out the 
ability of the algorithm 
to treat fermion zero modes in a 
clean and controllable manner.                

\end{abstract}                                                                  
\hspace{1cm}                                                                    
                                                                                
%%%%%%%%%%%%%%%%%%%%%%%%%%%%%%%%%%%%%%%%%%%%%%%%%%%%%%%%%%%%%%%%%%%%%%%%        
                                                                                
\pagebreak
%\section{   roduction}                         \label{sec:intro} 

In this letter we want to present a new algorithm for simulations
of dynamical fermions. 
Its basic conceptual idea is to separate out 
the low--lying eigenvalues 
of the Wilson-Dirac operator on the lattice and to 
not take 
this part of the spectrum
into account for the generation of the
gauge field configurations. However, the algorithm can be made exact
by incorporating the low--lying eigenvalues into the observables, or,
alternatively, by
adding them via a reject/accept step. 

From a principle point of view, 
the separation of the eigenvalue spectrum into a high and low frequency part 
allows 
to monitor the low--lying eigenvalues. In particular, 
the algorithm offers the possibility to detect 
the appearance of
eventual zero modes and to control their effects on physical observables. 
The low--lying eigenvalues are also expected to play an important role
in practice as 
they slow down the fermion simulation algorithms, when approaching
the chiral limit. Cutting these modes off, should therefore 
result in a gain for the cost of a practical simulation. 

The basic building blocks of the algorithm are the 
standard HMC algorithm \cite{hmc} 
and the multiboson technique to simulate dynamical fermions
\cite{luscher}. A similar idea has been presented shortly in 
\cite{takaishi}. 
In the multiboson technique, the inverse fermion matrix is
approximated by a polynomial written in powers
of the fermion matrix. 
We propose to take this polynomial to define the --approximate-- 
interaction of the fermions.

To be specific, let us consider the path integral
for 
%{\bf the $SU(3)$ gauge theory with} 
Wilson fermions on the lattice
\be \label{pf} 
{\mathcal Z}=\int {\mathcal D}U
       \exp\;\left\{- S_g\right\} \mbox{det}(Q^2)
 =\int {\mathcal D}U{\mathcal D}\phd{\mathcal D}\phi
       \exp\;\left\{- S_g-\phd Q^{-2}\phi \right\} \;\;.
\ee
The term $S_g$ in the exponential is the pure gauge action and is given by
\be
\label{eq:gauge}
S_g = -{\beta \over 6} \sum_{P} Tr(U_P+U^{\dagger}_P)\;\;.
\ee
The symbol $U_P$ represents the usual plaquette term on the lattice with
gauge links taken from SU(3). 
The determinant factor $\mbox{det}(Q^2)$ accounts for the contribution of
virtual fermion loops to the path integral.
The bosonic fields $\phi$ carry spinor, flavour and colour indices.
In eq.(\ref{pf}) and in the following we are assuming that we
have two mass-degenerate flavours.
The matrix $Q$ that appears in the determinant is a hermitian
 sparse matrix defined by:
\be
Q(U)_{x,y}=
c_0\gamma_5[\delta_{x,y}-
\kappa \sum_{\mu}(1-\gamma_{\mu})U_{x,\mu}\delta_{x+\mu,y}
+(1+\gamma_{\mu})U^{\dagger}_{x-\mu,\mu}\delta_{x-\mu,y}]\;\;,
\ee
with $\kappa$  the so-called hopping parameter, related to the bare
quark mass $m_0$ by $\kappa=(8+2m_0)^{-1}$, and
$c_0=[c_M (1+8\kappa)]^{-1}$, where $c_M$ should be chosen such that the
eigenvalues $\lambda$ of $Q$ satisfy $|\lambda |<1$. 

Let us assume that we have constructed a polynomial $P_{n}$ 
of degree $n$ 
such that
%, {\bf if all the eigenvalues of $Q^2$ are in the range $[\epsilon ,
% 1]$ } 
\begin{equation}
\mbox{det}\left[Q^2P_{n}(Q^2)\right] \rightarrow 1 \;\;\mbox{for}\;\; n\rightarrow \infty\; , 
\end{equation} 
with $P_{n}(\lambda(Q^2))>0$ for all the eigenvalues $\lambda(Q^2)$ of $Q^2$ in the 
range $0 \le \lambda(Q^2) < 1$.  
Then we can rewrite the 
determinant,
\be \label{determinant}
\mbox{det}(Q^2) = 
\frac{\mbox{det}\left[Q^2P_{n}(Q^2)\right]}
{\mbox{det}\left[P_{n}(Q^2)\right]}\; .
\ee 
Each of the two determinants on the right--hand side can be represented as a 
Gaussian integral 
with the help of bosonic
fields $\phi$ and $\eta$, respectively. 
The 
partition function becomes 
\be \label{newpf}
{\mathcal Z}
 =\int {\mathcal D}U{\mathcal D}\phd{\mathcal D}\phi {\mathcal D}\eta^\dagger{\mathcal D}\eta
      W \exp\;\left\{- S_g-\phd P_{n}(Q^2) \phi 
                   -\eta^\dagger\eta \right\}
\ee
where we introduced the ``correction factor'' 
\begin{equation} \label{correction} 
%W\equiv e^{w}\; ; \;\;
W=\exp\left\{\eta^\dagger\left(1-\left[Q^2P_{n}(Q^2)\right]^{-1}\right)\eta\right\} \;\;.
\end{equation}
Note that eq.(\ref{newpf}) is an exact rewriting of the partition function
eq.(\ref{pf}).

With the introduction of the correction factor $W$ 
the expectation value of an observable $O$ 
is now computed as 
\be \label{observable} 
< O > = \frac{<OW>_P}{<W>_P}\; , 
\ee
where the averages $<\dots >_P$ are taken with respect to the measure
%given in eq.(\ref{newpf}). 
%\be \label{polmeas}
%       {\mathcal D}U{\mathcal D}\phd{\mathcal D}\phi 
%       \exp\;\left\{- S_g-\phd P_{[n,\epsilon]}(Q^2) \phi \right\} 
%\ee
defined through the approximate fermion action
 $\phd P_{n}(Q^2) \phi$ .   
Alternatively one may incorporate the $W$ factor via a reject/accept
step. 

Let us now specify the form of the polynomial 
that we are going to take. We choose a 
Chebyshev polynomial to approximate $Q^{-2}$. 
When written in its factorized form 
\be \label{factorize}
P_{[n,\epsilon]}(Q^2) = c_N \prod_{k=1}^n \left[ Q^2 - z_k \right]
  = c_N \prod_{k=1}^n \left[ (Q - \sqrt{z_k}^{*})(Q - \sqrt{z_k}) \right]\; ,
\ee
it is characterized by its roots (for $k=1,2,\dots ,n$), 
%\be
%\mu_k + i\nu_k = \sqrt{z_k},\;\; \nu_k > 0.
%\ee
%with 
\be \label{roots}
z_k = \frac{1}{2}(1+\epsilon) - \frac{1}{2}(1+\epsilon)\cos(\frac{2\pi k}{n+1})
    - i\sqrt{\epsilon}\sin(\frac{2\pi k}{n+1})\;  
\ee
and a normalization factor $c_N$, which is explicitly calculable \cite{bunk}.
%The same polynomial may also be written in a different factorized form,
%as a polynomial of $Q$ (instead of $Q^2$) with roots given by $\sqrt{z_k}$
%and $\sqrt{z_k}^{*}$. 
The polynomial $P_{[n,\epsilon]}(s)$ approximates the function $1/s$ (where 
$s$ may correspond to any of the eigenvalues of $Q^2$)  
uniformly in the interval $\epsilon \le s \le 1$. 
%the case of practical 
%interest, considered in eq.(\ref{factorize}), is of course $s=Q^2$. 
The relative fit error 
\be
R_{[n,\epsilon]}(s) = \left[ P_{[n,\epsilon]}(s)-1/s \right] s
\ee
in this interval is exponentially small:
\be
|R_{[n,\epsilon]}(s)| \le 2 \left(\frac{1-\sqrt{\epsilon}}{1+\sqrt{\epsilon}}\right)^{n+1}
\; .
\ee
Let us finally 
introduce an accuracy parameter $\delta$, which is actually an upper bound
to the maximum relative error of the polynomial approximation, 
\be \label{accuracy}
\delta =  
2 \left(\frac{1-\sqrt{\epsilon}}{1+\sqrt{\epsilon}}\right)^{n+1} \; .
\ee
The parameter $\delta$ provides an easily computable and conservative 
measure of how well the chosen polynomial approximates $1/s$
in the given interval $\epsilon \le s \le 1$.

With the specification of the polynomial eq.(\ref{factorize}) the path
integral eq.(\ref{newpf}) and the correction factor $W$ are fully determined. 
It is clear that 
for polynomials of high degree, the interaction defined by them becomes too complicated
for the application of local algorithms. 
It is therefore a natural choice to use molecular dynamics 
algorithms like the HMC or the Kramers equation algorithms 
\cite{horowitz, kramers}. 
In the following we will call our hybrid of molecular dynamics and
multiboson algorithms 
the Polynomial Hybrid Monte Carlo (PHMC) algorithm. 

What do we expect from the PHMC algorithm? 
It has been suggested \cite{karlreview} that the lowest eigenvalue
of $Q^2$, $\lambda_{min}(Q^2)$, is an important quantity in 
determining the cost of the HMC algorithm. 
In particular, a theoretical analysis leads to the --optimistic-- 
estimate that the cost
grows as $1/\lambda_{min}^{3/2}(Q^2)$. In the PHMC algorithm, the role of the
lowest eigenvalue is taken over by the infrared cut-off parameter
$\epsilon$. 
For $0 \le s \le \epsilon$
the polynomial $P_{[n,\epsilon]}(s)$ is always finite with values $O(1/\epsilon)$. 
From the experience with the multiboson technique
\cite{bunk,beat,forcrand,galli,kramersboson} it has become clear 
that one might choose $\epsilon$ to be substantially larger than
$\lambda_{min}(Q^2)$ while still getting values for expectation values 
that are compatible with the ones obtained by the HMC algorithm. 
%Amazingly enough, {\bf in some cases and within typical statistical 
%errors,} this even holds for the lowest eigenvalue itself. 
This result suggests that one might choose $\epsilon > \lambda_{min}(Q^2)$ 
also in the PHMC algorithm without introducing too large fluctuations 
of the correction factor. Since in the PHMC algorithm only one 
bosonic field is introduced, one will also avoid the dangerous increase
of the autocorrelation time with increasing degree of the polynomial
as found for the multiboson technique \cite{beat}. 

Before we turn to the results
for the performance of the PHMC algorithm, 
let us shortly sketch, how the algorithm is implemented in our 
simulation program. 
We will be quite short here und refer to a 
forthcoming publication for more details and safety measures, in
particular when using the algorithm on a 32-bit arithmetics machine.
Let us start by discussing the heatbath for the bosonic fields
$\phi$, 
the action of which is given by
\be \label{bosonaction}
S_b=\phd P_{[n,\epsilon]}(Q^2) \phi\; . 
\ee
To generate a Gaussian distribution according to this interaction,
we proceed as follows. We first generate a Gaussian random vector $\zeta$. 
We then solve $Q^2P_{[n,\epsilon]}(Q^2)X = Q^2 \zeta$ using
a Conjugate Gradient (CG) method.
By writing $P_{[n,\epsilon]}(Q^2) = P^{*}_{n/2}(Q^2) P_{n/2}(Q^2)$ 
with appropriate ordering of 
the roots, we finally construct the $\phi$-fields
via $\phi = P_{n/2}^{*}(Q^2) X$. 

The derivation of the force for the PHMC algorithm is done in complete
analogy to the method used for the HMC algorithm \cite{gottlieb}. 
A variation of the action eq.(\ref{bosonaction}) using the polynomial 
eq.(\ref{factorize}) reveals that one has to construct the vectors 
\be \label{phij} 
\phi_j = \prod_{k=1}^{j}\left[Q-\sqrt{z_k}\right]\phi_0\;\; ; j=1,...,2n-1 
\ee
with $\phi_0$ the bosonic field generated by the boson heat bath. 
The vectors $\phi_j$ are precalculated and stored. This
calculation may be organized in such a way that the memory storage
required amounts to only $n+1$ (instead of $2n$) vectors $\phi_j$.   
One may then use them in the actual force computation at the
appropriate places. 

%Note that the degree $n$ of the polynomial plays now the role of the 
%number of iterations the CG method needs in the standard HMC algorithm. 
The above storage requirement for the vectors $\phi_j$
may be further 
reduced by introducing more bosonic field copies. For example, one
may split the polynomial into two parts $P_{n/2}^{(1)}$ and $P_{n/2}^{(2)}$
satisfying 
$det\left[ P_{[n,\epsilon]}(Q^2)\right]=det\left[ P_{n/2}^{(1)}(Q^2)\right]
\cdot det\left[ P_{n/2}^{(2)}(Q^2)\right]$ 
and integrate each contribution separately, leading to the action
\be
S_b=\phd_1 P_{n/2}^{(1)}(Q^2) \phi_1 + \phd_2 P_{n/2}^{(2)}(Q^2) \phi_2\; . 
\ee
%Although the theory obtained after integrating out the bosonic fields
%is not affected by this modification, the molecular dynamics behaviour
%(like the acceptance for a given trajectory) and eventually the autocorrelation
%times may be substantially changed. For this option, in any case, we did
%not yet perform a detailed analysis. }
It is amusing to note that by iterating this procedure one can obtain an interpolation 
between a nearly exact HMC algorithm, if $n$ and $1/\epsilon$ are large enough, and the
multiboson technique to simulate dynamical fermions. 
Although we did not yet perform an extended analysis, our first results 
indicate that introducing a few bosonic field copies does not increase the
autocorrelation time, when the number of copies is held small, say less 
than about eight.
It remains a subject of further study, however, whether the molecular dynamics 
behaviour is severely altered by the introduction of more field copies. 

Finally, the computation of the correction factor $W$ eq.(\ref{correction}) 
needs an additional
inversion of $Q^2P_{[n,\epsilon]}(Q^2)$. Since the $\eta$-field
occurring in $W$ is completely
independent from the $\phi$-field in the boson heatbath, 
this inversion has to be done separately. 

%In principle, the number of updates for the gauge fields and the 
%$\eta$-fields can be different and does not have to come in a 
%ratio of one. 
%
We decided to test the PHMC against the HMC algorithm on $4^4$ and 
$8^4$ lattices. All numerical results have been obtained on
Alenia Quadrics (APE) massively parallel computers. 
We adopted Schr\"odinger functional boundary conditions.
For the $4^4$ lattice we ran at $\beta = 6.4$, $\kappa= 0.15$ and for
the $8^4$ lattice we had $\beta = 5.6$ and 
$\kappa= 0.1585\approx \kappa_c$ \cite{gupta}. 
In both, the HMC and the PHMC algorithms, even-odd preconditioning 
\cite{precond}
and a Sexton-Weingarten leap-frog integration scheme \cite{sexy} 
is implemented. We want to emphasize that most of the improvements
to accelerate the HMC algorithm can be taken over to the PHMC algorithm. 
We think, therefore, that the results of the comparison we are performing here 
should be independent of the particular implementation. 

\begin{table*}[hbt]
% space before first and after last column: 1.5pc
% space between columns: 3.0pc (twice the above)
\setlength{\tabcolsep}{1.5pc}
% -----------------------------------------------------
% adapted from TeX book, p. 241
%\newlength{\digitwidth} \settowidth{\digitwidth}{\rm 0}
%\catcode`?=\active \def?{\kern\digitwidth}
% -----------------------------------------------------
\caption{Technical parameters for both algorithms}
\vspace{2mm}
\label{tab:table1}
\begin{tabular*}
%{\textwidth}{@{}l||@{\extracolsep{\fill}}cl|l@{\hspace{0.5cm}}l@{\hspace{0.5cm}}l@{\hspace{0.5cm}}l@{\hspace{0.5cm}}l@{\hspace{0.5cm}}}
{\textwidth}{@{}l||@{}cl|l@{\hspace{1.3cm}}l@{\hspace{1.3cm}}l@{\hspace{1.3cm}}l@{\hspace{1.2cm}}l@{\hspace{1.2cm}}}
\hline
                 & \multicolumn{2}{c|}{HMC} 
                 & \multicolumn{5}{c}{PHMC} \\
\cline{2-4} \cline{4-8}
Lattice & \makebox[1.7cm][c]{$\epsilon_{md}$} & $N_{md}$ & $\epsilon_{md}$ & $N_{md}$
        & $\epsilon$      & $n$ & $c_M$\\
\hline \hline
 $4^4$ & $0.25 $ & $4$  & $0.25 $  & $4$  & $0.036$ & $12$ & $0.5789$ \\
 $8^4$ & $0.075$ & $13$ & $0.09$ & $10$ &  $0.0026$  & $48$ & $0.5789$ \\
\hline
\end{tabular*}
\end{table*}

In table~1, we give the parameters of the algorithms which are the step size
$\epsilon_{md}$ and the number of molecular dynamics steps
$N_{md}$ as used for the leap frog integration. 
We also give the parameters characterizing the polynomial. 
The parameters were tuned in such a way that about the same 
acceptance rate was achieved in both algorithms, namely
82\% and 86\% for the HMC and PHMC algorithms, respectively, on the $4^4$ 
and, correspondingly, 
80\% and 79\% for the $8^4$ lattices.

%Note from table~\ref{tab:table1} that we could increase the stepsize 
%of the PHMC algorithm while maintaining the same acceptance as the
%HMC algorithm. 
With the choice of $c_M$ in table~\ref{tab:table1} we found  
the value of the largest eigenvalue 
$\lambda_{max}(\hat{Q}^2) $ of the preconditioned matrix 
$\hat{Q}$ to be close to 1.     
As first noted in \cite{borici}, 
since the polynomial is constructed such that it provides a uniform
approximation of $\hat{Q}^{-2}$ for $\epsilon < \lambda < 1$, 
lifting the eigenvalues by choosing
$c_M < 1$, allows to choose a larger value of 
$\epsilon$ and therefore a polynomial of lower degree
in order to achieve a desired value for the accuracy parameter $\delta$. 
%Of course, alternatively
%one might construct a polynomial that is optimized for the spectral 
%interval of the problem. 

We give in table~\ref{tab:table2} results for the expectation values of the plaquette 
$<P>$ and the lowest eigenvalue $<\lambda_{min}(\hat{Q}^2)>$ .
We also give
the uncorrected expectation values (setting $W=1$ in eq.(\ref{correction})) denoted
by a $*$. In the third column we give the number of trajectories
that were taken for the analysis. 

\begin{table*}[hbt]
% space before first and after last column: 1.5pc
% space between columns: 3.0pc (twice the above)
\setlength{\tabcolsep}{1.5pc}
% -----------------------------------------------------
% adapted from TeX book, p. 241
% -----------------------------------------------------
\caption{Results for both algorithms }
\vspace{2mm}
\label{tab:table2}
\begin{tabular*}{\textwidth}{@{}l@{\extracolsep{\fill}}lllll}
\hline
Lattice & Algorithm & $\#$ trajectories & $\langle P \rangle$ & $\langle \lambda_{min}(\hat{Q}^2) \rangle $ \\
\hline \hline
 $4^4$ & HMC          & $18000$ & $0.66179(13)$ & $0.01582(9)$  \\
       & PHMC         & $18000$ & $0.66169(16)$ & $0.01570(10)$  \\
       & PHMC$^{*}$   & $18000$ & $0.66248(13)$ & $0.01324(7)$  \\
 $8^4$ & HMC          & $2745$  & $0.57251(12)$ & $0.001310(51)$  \\
       & PHMC         & $2560$  & $0.57253(16)$ & $0.001328(51)$  \\
       & PHMC$^{*}$   & $2560$  & $0.57272(14)$ & $0.001141(51)$  \\
\hline
\end{tabular*}
\end{table*}

First of all, 
table~\ref{tab:table2} confirms the correctness of the PHMC algorithm.
While on the $4^4$ lattice the correction factor is important, 
one notices that for the $8^4$ lattice it only has a small effect. 
A crucial question is, whether the correction factor introduces 
strong fluctuations that may lead to large errors for the corrected
observables. We find that this is not the case, when we arrange for
a situation where $\epsilon$ is 2--3 times larger than the
lowest eigenvalue of the problem and the relative fit error of the
polynomial is kept small enough, $\delta \lsim 0.02$.
For larger values of $\delta$ the fluctuations can become
substantial, leading to large errors for the corrected observables
eq.(\ref{observable}).

%observables. We find that this is not the case, when {\bf
%the (by definition, uncorrected) average value of W, $<W>_P$, is not 
%too small (typically: $<W>_P \ge 0.1$). In practice, this may be achieved
%by choosing $\epsilon$ only few times larger than $c_M^2*\lambda_{min}(\
%hat{Q}^2$ and the degree $n$ of the polynomial such that $\delta$ is not too
%large (typically: $\delta \lsim 0.02$). A rough estimate of $<W>_P$,
%which allows to quickly check the choice of $n$ and $\epsilon$, is
%however obtained after few trajectories: this may be seen from the small errors
%on the values ($0.198(12)$ and $0.239(3)$ on $8^4$ and $4^4$ lattices, 
%respectively) that we get with the full statistics quoted in table~2.
%For larger values of $\epsilon$ and $\delta$, the value of $<W>_P$
%decreases and the fluctuations
%induced by the correction factor $W$ may become
%substantial, leading to large errors for the corrected observables, 
%as defined in eq.(\ref{observable}). }

In addition, the fluctuations of the corrected observables can be suppressed
further 
by choosing the number of updates of the $\eta$-fields to be larger than the 
number of full gauge field updates.
This amounts to compute the correction factor $N_{corr}$ times
on the same gauge field configuration. In our test, presented here, we
have chosen $N_{corr}=1$ on the $4^4$ lattice and $N_{corr}=2$ 
on the $8^4$ lattice.

%%%%%%%%%%%%%%%%%%%%%%% Figure for diagrams
\begin{figure}[t]
\vspace{-1mm}
\centerline{ \epsfysize=10.5cm
             \epsfxsize=10.5cm
             \epsfbox{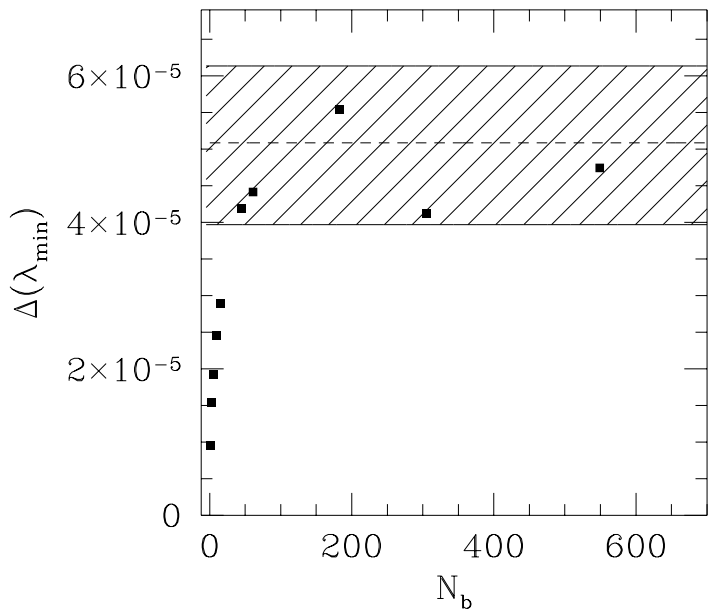}}
\begin{center}
\parbox{12.5cm}{\caption{ \label{fig:error}
The blocked jack-knife errors for the lowest eigenvalue $\lambda_{min}(\hat{Q}^2)$ 
from the
HMC algorithm (filled squares). $N_b$ is the binning block length. The dashed line
ist the estimate for the blocked error from the integrated autocorrelation time 
in the HMC algorithm. 
The shaded region is the estimate for the true error from the PHMC algorithm.
}}
\end{center}
\end{figure}
%%%%%%%%%%%%%%%%%%%%%%% Figure for diagrams

Since the behaviour of the observables from the HMC and the
corrected ones eq.(\ref{observable}) from the PHMC algorithm 
may in principle be very different, 
it is important to find an
estimate for the true error in order to be able to compare
both algorithms. To this end, we used a jack-knife 
binning procedure,
looking for a plateau in the blocked errors. For the HMC algorithm,
as a consistency check, we determined
also the integrated autocorrelation time computed directly from the autocorrelation
function. The result is illustrated in fig.~\ref{fig:error}
for the case of the lowest eigenvalue of $\hat{Q}^2$ on the $8^4$ lattice.
The filled squares are the blocked errors 
$\Delta(\lambda_{min}(\hat{Q}^2))$ as a function of the block length $N_b$
as obtained from the HMC algorithm.
$N_b =1$ corresponds to the naive error $\Delta_{naive}$, $N_b=2$ to blocking two 
consecutive measurements and so on. The dashed line indicates the true value of the
error as computed from the integrated autocorrelation time $\tau$, 
$\Delta_{true} = \sqrt{2\tau}\Delta_{naive}$.

For the PHMC algorithm on the $8^4$ lattice we ran  on the QH2 version
of the APE machine with 256 nodes. 
Distributing the $8^4$ lattice on 8 of these nodes, gives us 32 independent
systems, from which the error can be evaluated straightforwardly. 
One may also build from
these 32 systems 2 groups, each consisting of 16 independent systems  
and giving a separate error estimate $\Delta_1$ and $\Delta_2$. 
We take the difference between $\Delta_1$ and $\Delta_2$ 
as an estimate of the ``error of the error''.
We plot this uncertainty of the error 
as the shaded region in fig.~\ref{fig:error}. 
The same analysis can be made for the plaquette with a similar result. 
%{\bf Other analogous ways, that we also tried to estimate the ``error of
%the error'', gave similar and consistent results}

We conclude that with the same number of trajectories
both algorithms give  compatible
error bars for the plaquette and the lowest eigenvalue. Note, however, 
that with our statistics the error on the error is 
still significant. This is, of course, just a reflection of the uncertainty 
in the determination of the autocorrelation times. 

Let us discuss now the cost of a single trajectory in 
both algorithms. 
We write the total cost for the algorithms as
\be
C_{tot} = C_{Q\phi} + C_{extra} \; ,
\ee
where the first contribution is given by the number of matrix 
times vector $Q\phi$ operations and the second part accounts for all 
other operations. 
Asymptotically, when the condition number of $Q$ becomes
large, $C_{Q\phi}$ will by far dominate the cost of the algorithms. We will
therefore only discuss and compare the cost $C_{Q\phi}$ in
the following. Let us remark, however, that for small condition numbers 
$C_{extra}$ can be a non-negligible part of the total cost, in particular 
for the HMC algorithm as one might deduce from the details of the algorithm
structure. 
%{\bf(where more internal products per single
%$Q\phi$ operation occur)}. 

Let us denote by $N_{CG}$ the average number of iterations of the
Conjugate Gradient algorithm that is implemented in our
programs for all matrix inversions. Then the cost for the HMC algorithm 
in units of $Q\phi$ operations is given by
\be
C_{Q\phi}(HMC) = 2\cdot (2N_{md}+1)\cdot N_{CG}\; ,
\ee
where the first factor of 2 stems from the fact that one
needs 2 $Q\phi$ operations in each iteration of the CG routine. 
The factor $(2N_{md}+1)$ originates from the use of the 
Sexton-Weingarten integration scheme \cite{sexy}. 
The cost for the PHMC algorithm is split into three parts, 
\be
C_{Q\phi}(PHMC) = C_{bhb} + C_{update} + C_{corr}\; ,
\ee
where $C_{bhb}$ is the cost for the heatbath of the bosonic fields,
$C_{update}$ the cost for the force computation and $C_{corr}$ the cost 
to evaluate the correction factor. 
In units of $Q\phi$ operations we find 
\bea \label{PHMC_cost}
C_{bhb} & = & 2n\cdot  N_{CG}^{bhb} + n\nonumber \\ 
C_{update}  & = & 3n\cdot (2N_{md}+1)  \nonumber \\
C_{corr} & = & 2n\cdot  N_{CG}^{corr} \cdot  N_{corr}\; .
\eea
The factor $N_{corr}$ denotes as above the number of evaluations
of the correction factor $W$ per full gauge field update.
The factor $3n$ in $C_{update}$ comes for the following reason.
One needs basically $2n$ $Q\phi$ operations to construct the fields $\phi_j$ 
of eq.(\ref{phij}). The computation
of the total force needs a loop over the number of fields, $n$. In each iteration of
this loop one has to compute the variation of the action 
with respect to the gauge fields
for all four directions. This
computation corresponds roughly to one $Q\phi$ multiplication. We explicitly verified
this expectation for our implementation of the PHMC algorithm on the APE computer.
We expect, however, the formula for $C_{update}$ in the PHMC algorithm to also hold for
other situations like improved Wilson fermions. 

We give the cost of both algorithms in table~\ref{tab:table3}.
%The values of $N_{CG}$ used for the HMC cost in this table are $48$ 
%for the $4^4$ and $137$ for $8^4$ lattices. 
%For the PHMC case, the precise
%values of $N_{CG}^{bhb}$ and $N_{CG}^{corr}$ (typically 3 or 4) may be
%deduced from table~3, eq.(\ref{PHMC_cost}) and the values of $N_{corr}$
%mentioned above.}
We see that on the $8^4$ lattice, we win about a 
factor of $1.8$ against the HMC algorithm. 
Let us make a few remarks at this point. 

\begin{table*}[hbt]
% space before first and after last column: 1.5pc
% space between columns: 3.0pc (twice the above)
\setlength{\tabcolsep}{1.5pc}
% -----------------------------------------------------
% adapted from TeX book, p. 241
% -----------------------------------------------------
\caption{Cost for both algorithms }
\vspace{2mm}
\label{tab:table3}
\begin{tabular*}{\textwidth}{@{}l@{\extracolsep{\fill}}llllll}
\hline
Lattice & Algorithm & $C_{bhb}$ & $C_{update}$ & $C_{corr}$ & $C_{Q\phi}$ \\
\hline \hline
 $4^4$ & PHMC & 130    & 324    & 86    &  540  \\
       &  HMC & ---    & 868    & ---   & 868  \\
 $8^4$ & PHMC & $350$  & $3024$ & $600$ & $3974$  \\
       &  HMC & ---    & $7398$ &   --- & $7398$  \\
\hline
\end{tabular*}
\end{table*}

{\em (Correction factor)} 
We find for our PHMC algorithm that the plaquette has an autocorrelation
time of two while for the lowest eigenvalue the autocorrelation time
comes out to be about four. In this situation it is not necessary to
calculate the correction factor for every trajectory.
Indeed, the correction factor should be more considered
as part of the measurement and, from this point of view, its cost
should be added to the measurement costs. For observables, for which the
measurements are time consuming, or for situations where not every trajectory
is measured, the cost of the correction factor becomes negligible.

{\em (Long trajectories)} 
If we go to larger lattices and situations with larger
condition numbers than considered here, 
the number of steps per trajectory, $N_{md}$,
is increased when the trajectory length is kept fixed. 
We therefore expect that again the overheads for
the correction factor and also for the boson heatbath
will become negligible. From the above numbers, we conclude that the cost for the update
itself, $C_{update}$, is reduced in the PHMC algorithm 
by more than a factor of two as compared to standard HMC. 
%The above considerations also explain why we preferred to use 
%the HMC method instead of
%the Kramers equation method, which has by construction $N_{md}=1$. 

{\em (Parallelization)} 
When using massively parallel architectures, 
it is often advantageous to simulate several lattices 
simultaneously. If one uses the HMC algorithm, 
on SIMD architectures all the systems have
to wait until the system with the largest number of CG iterations has converged.
%{\bf If we denote with $N_CG^{max}$ the average (over all the trajectories)
%of the the largest number of CG iterations (among the several lattices),
%it is clear that $C_{Q\phi}(HMC)$ is proportional to $N_CG^{max}$ (instead 
%of $N_CG$), leading to an increase of the cost by a factor $N_CG^{max}/N_CG$.
%An estimate of this factor from our HMC data on the $8^4$ lattice with
%the physical parameters mentioned above gives about $1.4$.}  
If one compares the number of CG iterations from running 32 
replicas in parallel,
$N_{CG}({\rm 32 \;\; systems})$,  to
the one from running only a single system, $N_{CG}({\rm 1 \;\; system})$,
one finds that the ratio
$N_{CG}({\rm 32 \;\; systems})/N_{CG}({\rm 1 \;\; system})$
may easily reach values of about 2. 
In the PHMC algorithm, at least in the most time consuming part, the 
number of $Q\phi$ operations is, however, fixed by the degree of the
polynomial and the same for each system. 
We expect therefore for this situation the PHMC algorithm to give an
additional gain. Let us emphasize that, of course, all
of the costs of the algorithms given in table~3 refer to the
case of running a single system.

In conclusion, we have presented a new algorithm, called the PHMC algorithm
which is a hybrid of the standard HMC algorithm and the multiboson technique
to simulate dynamical fermions. Within the uncertainty of the error determination,
shown in fig.~{\ref{fig:error}, we find that for the same number of trajectories,
the errors from the HMC and the PHMC algorithms are 
about equal. At the same time,
the cost of generating a single trajectory is reduced by almost a factor of 2
when using the PHMC algorithm. Certainly, the properties of the PHMC algorithm 
have to be investigated more, different observables should be considered and,
of course, improved fermions should be studied. However, we find our results
very promising to finally find a real gain of about a factor of 2 over
the standard HMC algorithm. 

As a rule of thumb we advise to choose the lowest end of the fit range,
$\epsilon$, two or three times larger than the lowest eigenvalue of
$\hat{Q}^2$ and the degree $n$ of the fitting polynomial such that
the accuracy parameter $\delta$, introduced in eq.(\ref{accuracy}), is
between $0.01$ and $0.02$}. For too large values of $\delta$, the fluctuations
of the corrected observables eq.(\ref{observable}) become too large.
For too small values, the degree of the polynomial increases too much. 

Even more than the practical gain that we anticipate, we think 
that the PHMC algorithm has an advantage which is of principle nature. 
It has been demonstrated that for Wilson fermions, with and without
Symanzik improvement, fermionic 
(almost) zero modes may appear in the quenched approximation
\cite{eigenvalues,paperIII}. Such modes distort 
the statistical sample 
substantially. On the other hand, as discussed in \cite{paperIII},
the full path integral is finite and the fermion zero modes are cancelled
by the measure. 

The way the standard fermion simulation algorithms deal with the zero modes,
leaves us with a dilemma. Either these algorithms suppress the zero modes so
strongly that
in practical simulations
%, with a typical finite number of gauge field updates, 
configurations carrying (almost) zero modes do not occur at all. But then
we do not know what their importance is on physical observables, which is
unfortunate in particular within the context of topology. Or, on 
the other hand, a few configurations with (almost) fermion zero modes are
actually generated. But then they will lead to exceptional values for 
quark propagators and a reliable measurement of the observables involving
them will become very difficult.
%In any case, it would be very difficult 
%to understand the role of zero modes better, in particular within the
%context of topology.

In our PHMC algorithm, the update part is safe against the zero modes,
since the infrared cut-off parameter $\epsilon$ leaves the 
polynomial always finite. One may, however, monitor the lowest eigenvalue
and its eigenvector 
during a simulation by using minimization techniques like the one
described in \cite{kalk}. If an {\em isolated} 
zero mode is detected, one may switch from the
computation of the correction factor discussed above to the following
strategy. What we want to compute is 
\be
 det\left[ Q^2P(Q^2)\right] \equiv det\left[ A\right]\; .
\ee
Since we know the lowest eigenvalue 
$\lambda_{min}(A) = \lambda_{min}(Q^2)P(\lambda_{min}(Q^2))$ and its eigenvector
$\chi$, 
we may define a projector $P_\chi$ that projects onto the subspace 
orthogonal to $\chi$
%acting on a vector $\phi$ as 
%$P_1\phi = \phi - \chi(\chi,\phi)
leading to a matrix, where the lowest eigenvalue is taken out,
\be
\tilde{A} = A - \lambda_{min}(A) P_\chi\; .
\ee
Now, it is not difficult to show that
\be
det\left[ A\right] = \lambda_{min}(A)det\left[ \tilde{A}\right]\; , 
\ee
 where the factor $det\left[ \tilde{A}\right]$ may again be evaluated with the
help of Gaussian bosonic fields $\tilde{\eta}$.
%usual technique through a correction factor $\tilde{W}$ (defined analogously
%to $W$, but replacing $A$ with $\tilde{A}$).}

For pure gauge observables, like Wilson loops, we find therefore that
the configurations carrying zero modes have a negligible weight in 
eq.(\ref{observable}).
For a fermionic observable involving quark propagators, the situation 
is different because in the numerator of
eq.(\ref{observable}) the zero mode configurations may give a finite,
non--vanishing contribution, while 
in the denominator these configurations do not contribute. 
In this case the strategy will be to again
separate out the leading divergent contribution to the observable, 
which, when considering two degenerate quark flavours, may be at most
proportional to $\lambda_{min}^{-1}(Q^2)$.
Technically, this can be achieved again by projecting the lowest
eigenvalue out. The divergence possibly appearing in the leading contribution
will now be cancelled by the infinitesimal factor (proportional to
$\lambda_{min}(Q^2)$) from the correction factor yielding, as expected,
a finite, non--vanishing, well defined result. The non--leading contributions
to the fermionic observable ({\em i.e.} the ones less divergent than
$\lambda_{min}^{-1}(Q^2)$) may also be evaluated, by basically inverting
the matrix $Q^2 - \lambda_{min}(Q^2) P_\chi $, which is now well conditioned. 
%in the subspace orthogonal to $\chi$ the much better conditioned operator
%$Q^2 - \lambda_{min}(Q^2) P_\chi $. 
Note that these contributions are 
suppressed by the correction factor as $\lambda_{min}(Q^2) \to 0$.

The above discussion may be generalized to a situation where a number
of eigenvalues assume very small values. We therefore find that our PHMC 
algorithm is in principle able to take eventual zero modes into account in a controllable
way when performing dynamical fermion simulations. 

\vspace{0.5cm}
{\large\bf Acknowledgements}
%\vspace{3pt}
 
\noindent
 
This work is part of the ALPHA collaboration research program.
We thank M.~L\"uscher for essential comments and very helpful
discussions, in particular with respect to the role 
of zero modes in fermion simulation algorithms. 
All numerical simulation results have been obtained on the Alenia
Quadrics (APE) computers at DESY-IFH (Zeuthen). We thank the staff
of the computer center at Zeuthen for their support.
R.F. thanks the Alexander von Humboldt Foundation for
the financial support to his research stay at DESY.


\vspace{0.5cm}                                                                        
\begin{thebibliography}{99}                                                     
%
\bibitem{hmc}
S.~Duane, A.~D.~Kennedy, B.~J.~Pendleton and D.~Roweth, 
Phys. Lett. B195 (1987) 216.
%
\bibitem{luscher}
M.~L\"uscher, Nucl. Phys. B418 (1994) 637.
%
\bibitem{takaishi}
P. de Forcrand and T. Takaishi, hep-lat/9608093.
%
\bibitem{horowitz}
A.~M.~Horowitz, Phys. Lett. 156B (1985) 89; Nucl. Phys. B280 (1987) 510;
               Phys. Lett. 268B (1991) 247.
%
\bibitem{kramers} K. Jansen and C. Liu, Nucl. Phys. B453 (1995) 375; 
ibid. B459 (1996) 437.
%
\bibitem{karlreview}
K. Jansen, review talk at the International Symposium on Lattice Field
Theory, 1996, St. Louis, Mo, USA, hep-lat/9607051.
%
\bibitem{bunk} 
B.~Bunk, K.~Jansen, B.~Jegerlehner, M.~L\"uscher, 
H.~Simma and R.~Sommer, hep-lat/9411016; 
Nucl. Phys. B (Proc. Suppl.) 42 (1995) 49.
%
\bibitem{beat} B. Jegerlehner, Nucl.Phys.B465 (1996) 487.    
%
\bibitem{forcrand}
C. Alexandrou, A. Borrelli, P. de Forcrand, A. Galli and F. Jegerlehner,
Nucl.Phys.B456 (1995) 296.                       
%
\bibitem{galli}
A. Borrelli, P. de Forcrand and A. Galli, Nucl.Phys. B477 (1996) 809.
%
\bibitem{kramersboson}
K. Jansen, B. Jegerlehner and C. Liu, Phys.Lett. B375 (1996) 255. 
%
\bibitem{gottlieb} 
S.~Gottlieb, W.~Liu, D.~Toussaint, R.~L.~Renken and
R.~L.~Sugar, Phys. Rev. D 35 (1987) 2531.
%
\bibitem{gupta} R. Gupta et.al., Phys.Rev.D40 (1989) 2072.
%
\bibitem{precond}
T.~Degrand and P.~Rossi, Comp. Phys. Comm. 60  (1990) 211.
%
\bibitem{sexy} 
J.~C.~Sexton and D.~H.~Weingarten, Nucl. Phys. B380 (1992) 665.
%
\bibitem{eigenvalues}
K. Jansen, C. Liu, H. Simma and D. Smith, hep-lat/9608048.
%
\bibitem{paperIII}
M. L\"uscher, S.Sint, R. Sommer, P. Weisz and U. Wolff, CERN preprint,
CERN-TH/96-218, hep-lat/9609035.
%
\bibitem{kalk}
T, Kalkreuter and H. Simma, Comp.Phys.Comm. 93 (1996) 33. 
%
\bibitem{borici}
A.~Borici and P. de Forcrand, Nucl.Phys.B454 (1995) 645.
%
\end{thebibliography}
\end{document}